\documentclass[a4paper]{mem}
\usepackage{natbib}
\usepackage{txfonts}
\usepackage{balance}
\usepackage{graphicx}
\usepackage{dcolumn}
\newcolumntype{d}{D{.}{.}{-1}}
\def\dhead#1{\multicolumn{1}{c}{#1}}
\def\fns{\footnotesize}
\def\AD{{\sc ad}}
\def\fdg{.\!\!^\circ}
\def\itSigma{{\mathit\Sigma}}
\def\sigmaunit{W m$^{-2}$\,Hz$^{-1}$\,sr$^{-1}$}
\def\SNR(#1.#2)#3(#4.#5){{G#1${\cdot}$#2$#3$#4${\cdot}$#5}}
\def\HI{{H\,{\sc i}}}

\idline{01}{001}
\begin{document}

\title{Some statistics of Galactic SNRs}

\author{D.~A.\ Green}

\institute{Mullard Radio Astronomy Observatory, Cavendish Laboratory,
           Madingley Road,\\
           Cambridge CB3 0HE, United Kingdom\\
           e-mail: {\tt D.A.Green@mrao.cam.ac.uk}}

\authorrunning{Green}

\titlerunning{Galactic SNRs}

\abstract{The selection effects applicable to the identification of Galactic
supernova remnants (SNRs) at radio wavelengths are discussed. Low surface
brightness remnants are missing, as are those with small angular sizes
(including young but distant SNRs). Several statistical properties of Galactic
SNRs are discussed, including the surface-brightness/diameter ($\itSigma{-}D$)
relation. The wide range of intrinsic properties of Galactic remnants with
known distances, and the observational selection effects, means that the
$\itSigma{-}D$ relation is of limited use to derive diameters and hence
distances for individual SNRs, or for statistical studies.
}

\maketitle{}

\section{Introduction}

Over two hundred Supernova Remnants (SNRs) have been identified in our Galaxy.
I have produced several versions of a catalogue of Galactic SNRs over the last
twenty years, the most recent revised in 2004 January \citep{Gre04}. Here I
review some of the statistical properties of Galactic remnants based on the
most recent version of the catalogue. In particular I emphasize the importance
that the selection effects applicable to the identification of Galactic SNRs
must be appreciated.

\section{The catalogue}\label{s:catalogue}

The catalogue of Galactic SNRs exists in two formats. First -- published as an
appendix to \citet{Gre04} -- the basic parameters (Galactic and equatorial
coordinates, size, type, radio flux density, spectral index, and other names)
for each remnant. Second, a more detailed version, available on the
World-Wide-Web\footnote{See {\scriptsize\tt
http://www.mrao.cam.ac.uk/surveys/snrs/}.} which includes the descriptions of
each remnant, additional notes and references. The detailed version of the
catalogue is available as postscript or pdf for downloading and printing, or as
web pages. The web pages include links to the `NASA Astrophysics Data System'
for each of the nearly one thousand references. Notes both on those objects no
longer thought to be SNRs, and on the many possible and probable remnants that
have been reported, are also included in the detailed version of the catalogue.
It should be noted that the catalogue is far from homogeneous. It is
particularly difficult to be uniform in terms of which objects are considered
as definite remnants, and are included in the catalogue, rather than listed as
possible or probable remnants which require further observations to clarify
their nature. Since the first version of the catalogue was published in
\citet{Gre84} the number of identified Galactic SNRs has increased considerably, from
145 to 231. Much of this increase has been due to the availability of large
area radio surveys, particularly the Effelsberg survey at 2.7~GHz, and the MOST
survey at 843~MHz, which are discussed further below (Section \ref{s:surface}).

\section{Selection effects}\label{s:selection}

In practice the dominant selection effects applicable to identification of
Galactic SNRs are those that are applicable at radio wavelengths.
Simplistically, two selection effects apply \citep[{e.g.}][]{Gre91}, due to the
difficulty in identifying (i) faint remnants and (ii) small angular size
remnants.

\subsection{Surface brightness}\label{s:surface}

SNRs need to have a high enough surface brightness for them to be distinguished
from the background Galactic emission. This selection effect is {\em not}
uniform across the sky, both because the Galactic background varies with
position, and because the sensitivities of available wide area surveys covering
different portions of the Galactic plane vary. The most recent large-scale
radio surveys that have covered much of the Galactic plane are: (i) the
Effelsberg survey at 2.7~GHz \citep{Rei90,Fur90}, which
covered $358^\circ < l < 240^\circ$ and $|b| < 5^\circ$; and (ii) the MOST
survey at 843~MHz \citep{Whi96, Gre99}\footnote{To avoid confusion between authors with the same surname, I
include initials for Anne J.\ Green, of the University of Sydney, Australia.},
which covered $245^\circ < l < 355^\circ$, but only to $|b| < 1\fdg5$. In the
current catalogue of SNRs, fainter remnants are relatively more common in the
anti-centre and away from $b=0^\circ$, where the Galactic background is lower
\citep[see][]{Gre04}.

Since the new SNRs identified from the Effelsberg survey were included in the
version of the SNR catalogue published in \citet{Gre91}, the surface
brightnesses of remnants in the survey region that have subsequently been
identified are useful for estimating the completeness limit for this survey.
Since 1991 an additional 24 remnants within $358^\circ < l < 240^\circ$ and
$|b| < 5^\circ$ have been included in the catalogue, most in the first
quadrant. The surface brightnesses of these remnants suggest a completeness
limit of $\itSigma \approx 10^{-20}$ {\sigmaunit}, at 1~GHz, for the Effelsberg
survey \citep[see][for further discussion]{Gre04}. And since the number of
remnants brighter than this value is similar in the 1st and 4th quadrants, then
a similar limit seems appropriate for the MOST survey region.

\begin{figure}
\centerline{\includegraphics[width=66mm]{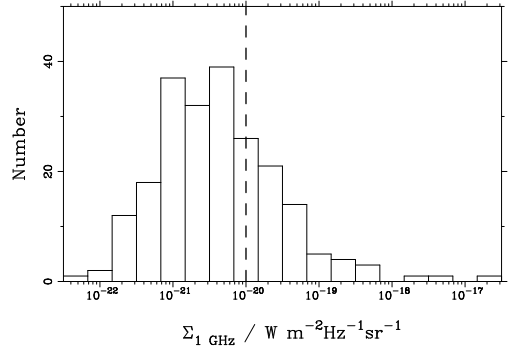}}
\caption{\footnotesize Distribution in surface brightness at 1~GHz of 217 Galactic SNRs. The
dashed line indicates the surface brightness completeness limit discussed in
Section~\ref{s:surface}.\label{f:sigma}}
\end{figure}

\begin{figure*}
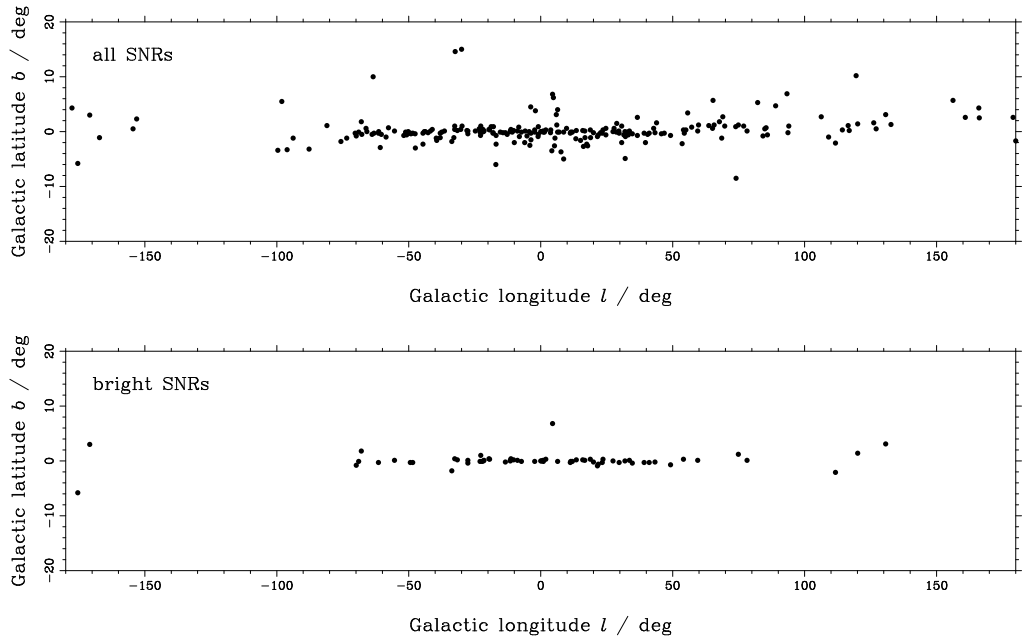

\centerline{\includegraphics[angle=270,width=13.5cm]{Green2a}}
\quad\\
\centerline{\includegraphics[angle=270,width=13.5cm]{Green2b}}
\caption{\footnotesize Galactic distribution of (top) all Galactic SNR and
(bottom) those SNRs with a surface brightness at 1~GHz greater than $10^{-20}$
{\sigmaunit}. (Note that the latitude and longitude axes are not to
scale.)\label{f:lb}}
\end{figure*}

Thus, the surface brightness limit for completeness of the current catalogue of
Galactic SNRs is approximately $10^{-20}$ {\sigmaunit}. Fig.~\ref{f:sigma}
shows a histogram of the surface brightnesses of the 217 Galactic SNRs, of
which 64 are above this nominal surface brightness limit.\footnote{For 14
catalogued remnants no reliable radio flux density, or only a limit is
available.} The SNRs with surface brightnesses below this limit are
predominantly in regions of the Galaxy where the background is low, i.e.\ in
the 2nd and 3rd quadrants, and away from $b=0^\circ$, as shown in
Fig.~\ref{f:lb}.

Ongoing and future observations will no doubt continue to detect more Galactic
SNRs, although it seems very likely that most of these objects will be faint,
and hence difficult to study in detail. Currently there are several large scale
radio surveys underway that will cover much of the Galactic plane
\footnote{See: {\scriptsize\tt http://www.ras.ucalgary.ca/IGPS/} for further
information.}. As is discussed in Section~\ref{s:sigmad}, there is a general
trend that fainter remnants tend to be larger, and hence on average older, than
brighter remnants. However, because of the wide range of properties of Galactic
SNRs with known distances, the surface brightness selection effect applies not
just to old remnants, but also to young remnants.

\begin{figure}
\centerline{\includegraphics[width=66mm]{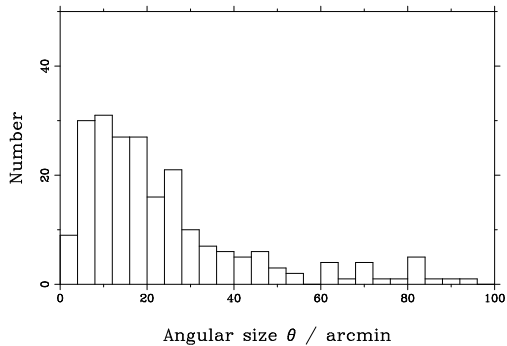}}
\caption{\footnotesize Histogram of the angular size of 219 Galactic SNRs (12
remnants larger than 100 arcmin are not included).\label{f:theta}}
\end{figure}

\subsection{Angular size}

Current catalogues are also likely to be missing small angular size SNRs, as
their structure is not well resolved by the available Galactic plane surveys,
and they would not be recognised as likely SNRs. Fig.~\ref{f:theta} is shows
the histogram of the angular sizes of known remnants, which peaks at around
10~arcmin. The limiting angular size varies for the different available wide
area surveys. As discussed above, the radio survey that covers most of the
Galactic plane is the Effelsberg 2.7-GHz survey, which has a resolution of
$\approx 4.3$~arcmin. So, for this survey, any remnants less than about
13~arcmin in diameter (i.e.\ 3 beamwidths) are not likely to be recognised from
their structures (although, as discussed below in Section~\ref{s:missing}, some
searches have been made for small remnants among compact sources in the
Effelsberg 2.7-GHz survey). The MOST 843-MHz survey has a much better
resolution, $\approx 0.7$~arcmin, which implies that in the region of the
Galactic plane covered by this survey only remnants smaller than about $\approx
2$ arcmin (i.e.\ 3 beamwidths) might be expected to be missed. However,
although the MOST survey detected 18 new SNRs \citep{Whi96},
the smallest new remnant was \SNR(345.7)-(0.2), which is $7 \times 5$
arcmin$^2$ in extent, i.e.\ several times larger than the nominal limit of
$\approx 2$ arcmin. Thus it is difficult to quote a single angular size
selection limit for current SNR catalogues, although it is clear that it is
difficult to identify small angular size remnants from existing wide area
surveys.

\subsection{Missing young but distant SNRs}\label{s:missing}

The lack of small angular size remnants is particularly clear when the remnants
of known `historical' Galactic supernovae \citep[see][]{Ste02} are considered.
These remnants are relatively close-by -- as is expected, since their parent
SNe were seen historically -- and therefore sample only a small fraction of the
Galactic disc. Consequently many more similar, but more distant remnants are
expected in our Galaxy \citep[{e.g.}][]{Gre85}, but these are not present in
the current catalogue.

\begin{table*}
\caption{\footnotesize Parameters of known historical SNRs, plus
Cas~A.\label{t:historical}}
\begin{center}
\small\tabcolsep 1.5pt
\begin{tabular}{cccdcdccdccd}\hline
\phantom{$\int^\Sigma$}
           &                    &          & \multicolumn{3}{c}{as observed}                     & & \multicolumn{2}{c}{if at 8.5~kpc} & & \multicolumn{2}{c}{if at 17~kpc} \\ \cline{4-6}\cline{8-9}\cline{11-12}
\phantom{$\int^\Sigma$}
    date   & name or            & distance &   \dhead{size}   & $\itSigma_{\rm 1~GHz}$ & \dhead{$S_{\rm 1~GHz}$}& &   size    & \dhead{$S_{\rm 1~GHz}$} & &   size    & \dhead{$S_{\rm 1~GHz}$} \\
           & remnant            &   /kpc   &  \dhead{/arcmin} &   {\fns /{\sigmaunit}} &       \dhead{/Jy}      & &  /arcmin  &      \dhead{/Jy}        & &  /arcmin  &       \dhead{/Jy}       \\ \hline
\phantom{$\int^\Sigma$}
     --    & Cas A              &   3.4    &     5     & $1.6 \times 10^{-17}$ &     2720        & &    2.0     &      435             & &    1.0    &   109                \\
{\AD} 1604 & Kepler's           &   2.9    &     3     & $3.2 \times 10^{-19}$ &       19        & &    1.0     &        2.2           & &    0.5    &     0.55             \\
{\AD} 1572 & Tycho's            &   2.3    &     8     & $1.3 \times 10^{-19}$ &       56        & &    2.3     &        4.1           & &    1.1    &     1.0              \\
{\AD} 1181 & 3C58               &   3.2    &     7     & $1.0 \times 10^{-19}$ &       33        & &    2.6     &        4.7           & &    1.3    &     1.2              \\
{\AD} 1054 & {\fns Crab nebula} &   1.9    &     6     & $4.4 \times 10^{-18}$ &     1040        & &    1.4     &       52             & &    0.7    &    13                \\
{\AD} 1006 & {\fns \SNR(327.6)+(14.6)} &   2.2    &    30     & $3.2 \times 10^{-21}$ &       19        & &    7.7     &        1.3           & &    3.9    &     0.31             \\ \hline
\end{tabular}
\end{center}
\end{table*}

Table~\ref{t:historical} gives the distances, angular sizes, flux densities and
surface brightnesses at 1~GHz, for the remnants of known historical supernovae
from the last thousand years, plus Cas~A (which although its supernova was not
seen -- so it is not strictly a historical remnant -- is known to be only about
300 years old). This table also lists the parameters of these remnants when
scaled to larger distances of 8.5 and 17~kpc, i.e.\ to represent how they would
appear if they if they were at the other side of the Galaxy.

From Table~\ref{t:historical}, any young SNRs in the Galaxy similar to the
known historical remnants, but in the far half of the Galaxy, would generally
be expected to have angular sizes less than a few arcmin, usually with high
surface brightness, greater than $\approx 10^{-19}$ {\sigmaunit} (although
remnants similar to the remnant of the SN of {\AD} 1006 would be much fainter).
These remnants would also be expected to lie close to the Galactic plane, with
$|b| \lesssim 1^\circ$. Several dozen other young (i.e.\ less than a thousand
year old) SNRs are expected in the Galaxy \citep[see][for further
discussion]{Gre04}. However,  there are very few such remnants in the current Galactic
SNR catalogue. In fact there are only 3 known remnants with angular sizes of 2
arcmin or less: \SNR(1.9)+(0.3), \SNR(54.1)+(0.3) and \SNR(337.0)-(0.1). (It
should be noted that there are unlikely to be other luminous remnants in the
Galaxy like Cas A and the Crab nebula. Any such remnants, even on the far side
of the Galaxy, would have relatively high flux densities, and the nature of all
such sources in the Galactic plane is known.)

Since the missing young but distant remnants are expected to have angular sizes
of a few arcmin or less, they will not have been resolved sufficiently in large
area radio surveys. Higher resolution, targeted observations are needed to
identify any such small remnants, and although there have been several such
searches for remnants of this type ({e.g.} \citealt{Gre84+,
Hel85, Gre85, Gre89, Sra92, Mis02},
see also
\citealt{Sai04}), they have had only limited success (identifying the small
remnants \SNR(1.9)+(0.3) and \SNR(54.1)+(0.3) noted above). (An additional
candidate small, presumably young SNR is \SNR(337.2)+(0.1), which was listed as
a possible SNR by Whiteoak \& A.~J.\ Green. Recently \citet{Com05} --
see also these proceedings -- have also noted that this source is associated
with an X-ray source listed by \citet{Sug01}, which supports the SNR
identification.\footnote{Note that the radio surface brightness for
\SNR(337.2)+(0.1) given by \citet{Com05} is too low by two orders of
magnitude.})

The fact that such missing, small remnants are likely to be in complex regions
of the plane may mean that confusion is a very significant problem, and not
just at radio wavelengths. Further searches for these missing young but distant
remnants are clearly required.

\section{Some simple SNR statistics}\label{s:simple}

In the current version of the catalogue, 77\% of remnants are classed as shell
type (or possible shell), 12\% are composite (or possible composite), and 4\%
are filled-centre (or possible filled centre) remnants. The remaining 7\% have
not yet been observed well enough to be sure of their type, or else are objects
which are conventionally regarded as SNRs although they do not fit well into
any of the conventional types (e.g.\ CTB80 ($=$\SNR(69.0)+(2.7)), MSH
17$-$3{\em 9} ($=$\SNR(357.7)-(0.1))).

There are 14 Galactic SNRs that are either not detected at radio wavelengths,
or are poorly defined by current radio observations, so that their flux density
at 1~GHz cannot be determined with any confidence: i.e.\ 94\% have a flux
density at 1~GHz included in the catalogue. Of the catalogued remnants, 36\%
are detected in X-ray, and 23\% in the optical. At both these wavelengths,
Galactic absorption hampers the detection of distant remnants.

\section{Distance dependent SNR statistics}\label{s:distance}

\subsection{Distances to SNRs}\label{s:distances}

Accurate distances are not available for most known SNRs, which is a problem
for many studies of Galactic SNRs, where it is necessary to know the distances
to remnants (or equivalently their physical sizes, since their angular sizes
are known). The distances that are available are obtained from a wide variety
of methods, each of which is subject to their own uncertainties, and some of
which are subjective. In the next few years further distance measurements
should become available for more SNRs, both because of the various ongoing
Galactic plane survey, and new distance measurement techniques: e.g.\ from
{\HI} column densities \citep[see][]{Fos03}, or {\HI} absorption to
polarised emission from remnants \citep{Kot04}. Currently 
\cite[see][]{Gre04}, the distances to 47 Galactic SNRs 
are available, i.e.\ only 20\% of
known Galactic SNRs. The uncertainties in these distances are far from uniform.
For kinematic distances -- which are a large majority of the available
distances -- there are always some uncertainties in deriving distances from
observed velocities, due to deviations from circular motion (especially an
issue for nearby remnants, and for those near $l=0^\circ$ and $180^\circ$ where
the observed velocity does not depend strongly on distance) and ambiguities
inside the Solar Circle.

\subsection{The $\itSigma{-}D$ and $L{-}D$ Relations}\label{s:sigmad}

Many statistical studies of Galactic SNRs have relied on the
surface-brightness/diameter, or `$\itSigma{-}D$' relation to derive distances
for individual SNRs from their observed flux densities and angular sizes. For
remnants with known distances ($d$), and hence known diameters ($D$),
physically large SNRs are fainter (i.e.\ they have a lower surface brightness)
than small remnants. Using this correlation between $\itSigma$ and $D$ for
remnants with known distances, a physical diameter is deduced from the
distance-independent {\em observed} surface brightness of any remnant. Then a
distance to the remnant can be deduced from this diameter and the observed
angular size of the remnant.

\begin{figure}
\centerline{\includegraphics[angle=270,width=66mm]{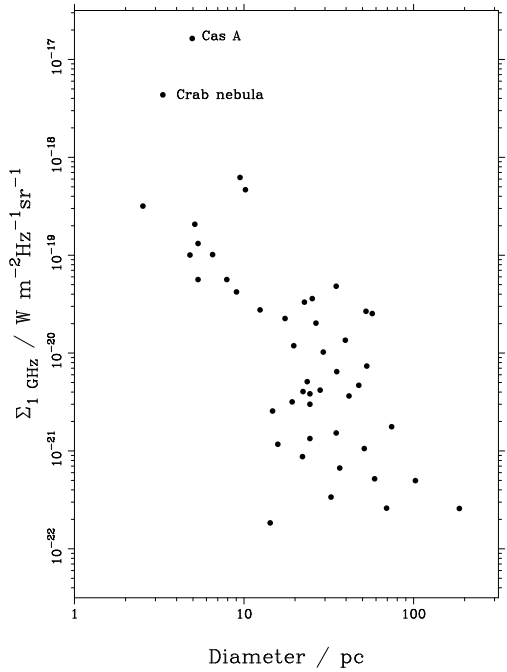}}
\caption{\footnotesize The surface brightness/diameter ($\itSigma{-}D$)
relation for 47 Galactic SNRs with known distances. Note that the lower left
part of this diagram is likely to be seriously affected by selection
effects.\label{f:sigmad}}
\end{figure}

The $\itSigma{-}D$ relation for Galactic SNRs with known distances is shown in
Fig.~\ref{f:sigmad}. This clearly shows a wide range of diameters for a given
surface brightness, which is a severe limitation in the usefulness of the
$\itSigma{-}D$ relation for deriving the diameters, and hence distances, to
individual remnants. For a particular surface brightness, the diameters of SNRs
vary by up to about an order of magnitude, or conversely, for a particular
diameter, the range of observed surface brightnesses seen varies by more than
two orders of magnitude.

\begin{figure}
\centerline{\includegraphics[angle=270,width=66mm]{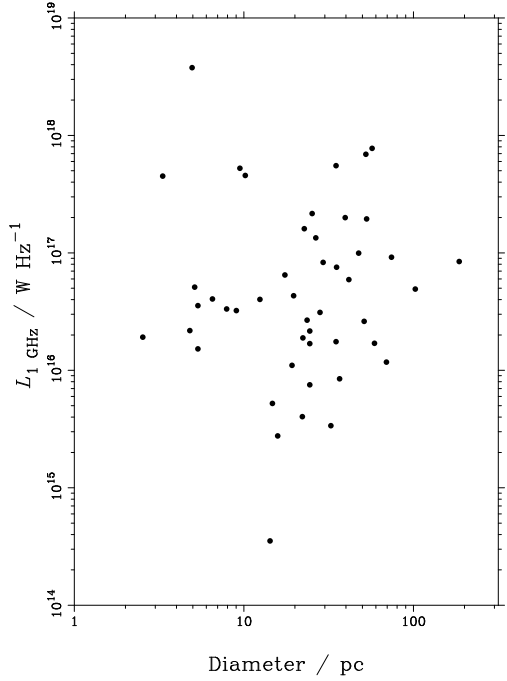}}
\caption{\footnotesize The luminosity/diameter ($L{-}D$) relation for 47
Galactic SNRs with known distances.\label{f:ld}}
\end{figure}

The correlation shown between surface brightness and diameter in
Fig.~\ref{f:sigmad} is, however, largely a consequence of the fact that it is a
plot of surface-brightness -- rather than luminosity -- against diameter, $D$.
Surface brightness is plotted, because it is the distance-independent
observable that is available for (almost) all SNRs, including those for which
distances are not available. For remnants whose distances are known, we can
instead consider the radio luminosity of the remnants. Since $\itSigma$ and
luminosity, $L$, depends on the flux density $S$, angular size $\theta$,
distance $d$ and diameter $D$, as
$$
  \itSigma \propto {S \over \theta^2} \quad{\rm and}\quad
  L \propto S d^2
$$
then
$$
   \itSigma \propto {L \over (\theta d)^2} \quad{\rm or}\quad
   \itSigma \propto {L \over D^2}.
$$
Thus, much of the correlation shown in the $\itSigma{-}D$ relation in
Fig.~\ref{f:sigmad} is due to the $D^{-2}$ bias that is inherent when plotting
$\itSigma$ against $D$, instead of $L$ against $D$. The $L{-}D$ relation for
Galactic SNRs with known distances in Fig.~\ref{f:ld} shows that there is wide
range of luminosities for SNRs of all diameters. Cas~A is the most luminous
Galactic SNR, but it appears to be at the edge of a wide distribution of
luminosities. The wide range of luminosities is perhaps not surprising, given
that the remnants are produced for a variety of types of supernovae, and that
they evolve in regions of ISM with a range of properties (e.g.\ density), which
may well effect the efficiency of the radio emission mechanism at work. For
example, some SNRs may initially evolve inside a low-density, wind-blown
cavity, and then collide with the much denser regions of the surrounding ISM.

Furthermore, the full range of intrinsic properties of SNRs may be even wider
than that shown in Figs~\ref{f:sigmad} and \ref{f:ld}, as the selection effects
discussed in Section \ref{s:selection} mean that it is difficult to identify
small and/or faint SNRs.

Also, it is not clear that any best-fit $\itSigma{-}D$ relation -- not
withstanding selection effect problems -- actually represents the evolutionary
track of individual SNRs (see, for example, \citealt{Ber86}, or
\citealt{Ber04} for a recent discussion). The distribution of SNRs with known
distances is a snapshot in time of a population of remnants, and individual
remnants may evolve in the $\itSigma{-}D$ plane in directions quite different
from any power law fitted to the overall distribution of SNRs (or to the upper
limit of the distribution). As a simple example, consider the situation where
SNRs have a range of intrinsic luminosities, expand with a constant luminosity
up to some particular diameter --  which varies for different SNRs depending on
their environment (e.g.\ the surrounding ISM density, which influences their
expansion speed, which may affect the efficiency of radio emission mechanism)
-- after which their radio luminosity fades rapidly. In this case, the locus of
the upper bound to the highest surface brightness remnants for a particular
diameter is related to where the luminosities of different SNRs begin to
decrease, and does not represent the evolutionary track of any individual
remnant.

\subsection{A question of regression}

There is a further issue related to any power law $\itSigma \propto D^{n}$ fits
to the observed properties of samples of SNRs, even if any problems with the
selection effects are neglected. The $\itSigma{-}D$ relation has two particular
uses, first, the derivation of diameters (and hence distances) for Galactic
remnants from their observed surface brightnesses, and second, to parameterise
the relationship between surface brightness and diameter, for comparison with
models and theories, or between different samples of SNRs. For the former, the
measured value of $\itSigma$ is used to predict $D$, and in this case a least
squares fit minimizing the deviations in $\log D$ should be used (see
\citealt{Iso90} for a discussion of least squares fitting). This, however, has not
always been done, and instead fits minimizing the deviations in $\log \itSigma$
have been used. The differences between these two least square fits can be very
large for remnants much fainter or brighter than the average surface of the
remnants with known distances used to calibrate the $\itSigma{-}D$ relation.
For example, \citet{Cas98} derived a best fit $\itSigma{-}D$
relation with a slope ($n$ above) of $-2.38 \pm 0.26$ using a sample of 36
shell SNRs (excluding Cas A) evidently by minimizing the square of the
deviations in $\log \itSigma$. Instead minimizing the deviations in $\log D$ --
which is appropriate if the relation is to be used to {\em predict} diameters
from surface brightnesses -- leads to a much steeper fit, with a power law
slope of $-3.37 \pm 0.35$. This means that Case \& Bhattacharya's ``best fit''
overestimates the diameters, and hence also the distances of fainter remnants.
To quantify this, a faint remnant with a surface brightness of $3 \times
10^{-22}$ {\sigmaunit} at 1~GHz, would have its diameter overestimated by about
40\%, which is larger than the nominal uncertainty that Case \& Bhattacharya
claimed (although there is evidently a wide spread of intrinsic properties of
SNRs with known distance, see Fig.~\ref{f:sigmad}). If a $\itSigma \propto
D^{n}$ relation is to be used to describe the relationship between $\itSigma$
and $D$, then arguably (see Isobe et al.) a fit to observations that treats
$\itSigma$ and $D$ symmetrically is appropriate (e.g.\ the bisector of least
square fits made minimizing either $\log \itSigma$ or $\log D$).

\section{Galactic SNR distribution}\label{s:distribution}

The distribution of SNRs in the Galaxy is of interest for many astrophysical
studies, particularly in relation to their energy input into the ISM and for
comparison with the distributions of possible progenitor populations. Such
studies are, however, not straightforward, due to observational selection
effects and the lack of reliable distance estimates available for most
identified remnants. In particular, all SNRs in the anti-centre (i.e.\ 2nd and
3rd Galactic quadrants) are outside the Solar Circle, at large Galactocentric
radii. These are regions where the background Galactic emission is faint, so
that low surface brightness remnants are relatively easy to identify (see
Section~\ref{s:selection}). Without taking selection effects into account, the
larger number of fainter SNRs in the anti-centre leads to an apparently broad
distribution of Galactic SNRs in Galactocentric radius 
\citep[e.g.\ see further
discussion in][]{Gre04}. A simple approach is to compare the distribution in
Galactic longitude of bright remnants, with $\itSigma_{\rm 1~GHz} > 10^{-20}$
{\sigmaunit} where selection effects are not important, with the expected
distribution for various models. In \citet{Gre04} this was done with 64 bright
remnants, and simple Monte Carlo models of a Gaussian distribution, where the
probability distribution varies with Galactocentric radius, $R$, as $\propto
{\rm e}^{-(R/\sigma)^2}$, where $\sigma$ is the Galactocentric scale length. A
simple least squares comparison\footnote{Not a $\chi^2$ comparison, as
erroneously stated in \citet{Gre04}.} of the observed and model cumulative
distributions indicates that for this simple model a scale length of $\sigma
\approx 6.5$~kpc best matches the observed distribution of high brightness
SNRs.

\section{Conclusions}

In statistical studies of Galactic SNRs it is crucial that the selection
effects that apply to the identification of remnants are taken into account.
Both faint (i.e.\ low surface brightness) and small angular size remnants are
missing from current catalogues of Galactic SNRs. A consequence of the angular
size selection effect is that few young but distant remnants have yet been
identified in the Galaxy. For remnants with known distances, the intrinsic
range of luminosity of Galactic SNRs is large, which combined with selection
effects, means that the $\itSigma{-}D$ relation is of limited use for
determining distances to individual remnants, or for statistical studies.

\begin{acknowledgements}
I am grateful to many colleagues for numerous comments on, and corrections to,
the various versions of the Galactic SNR catalogue. This research has made use
of NASA's Astrophysics Data System Bibliographic Services.
\end{acknowledgements}

\bibliographystyle{aa}

\end{document}